\documentclass[twocolumn,secnumarabic,amssymb, nobibnotes, aps, prd]{revtex4-1}

\usepackage{amsmath, amssymb}
\usepackage{graphicx}
\usepackage{color}
\usepackage{stackengine,graphicx}
\usepackage{textcmds}
\usepackage{mathtools}
\usepackage{mathtools}
\usepackage{ulem}
\usepackage[utf8]{inputenc}

\begin{document}

\title{
       Fluctuation induced network patterns in active matter with spatially correlated noise 
       }

\author{Sebastian Fehlinger\textsuperscript{\dag,}\footnote{sebastian.fehlinger@pkm.tu-darmstadt.de}, Kai Cui\textsuperscript{\ddag}, Arooj Sajjad\textsuperscript{\S}, Heinz Koeppl\textsuperscript{\ddag}, and Benno Liebchen\textsuperscript{\dag,}\footnote{benno.liebchen@pkm.tu-darmstadt.de}}
\email[]{benno.liebchen@pkm.tu-darmstadt.de}
\affiliation{\textsuperscript{\dag}Technische Universität Darmstadt, Department of Physics, Institute of Condensed Matter Physics, Hochschulstrasse 8, 64289 Darmstadt, Germany}
\affiliation{\textsuperscript{\ddag}Technische Universität Darmstadt, Department of Electrical Engineering and Information Technology, Centre for Synthetic Biology,  Merckstrasse 25, 64283 Darmstadt, Germany}
\affiliation{\textsuperscript{\S}Technische Universität Darmstadt, Department of Biology, Poppinga lab: Biomechanics, Functional Morphology, and Biomimetics, Schnittspahnstrasse 11, 64287 Darmstadt, Germany}




\begin{abstract}
Fluctuations play a central role in many fields of physics, from quantum electrodynamics to statistical mechanics. In active matter physics, most models focus on thermal fluctuations due to a surrounding solvent. An alternative but much less explored noise source can occur due to fluctuating external fields, which typically feature certain spatial correlations. In this work, we introduce a minimal model to explore the influence of spatially correlated but temporally uncorrelated noise on the collective behavior of active particles. We find that specifically in chiral active particles such fluctuations induce the formation of network patterns, which neither occur for spatially (uncorrelated) thermal noise, nor in the complete absence of fluctuations. These networks show (i) a percolated structure, (ii) local alignment of the contained particles, but no global alignment, and (iii) hardly coarsen. We perform a topological data analysis to systematically characterize the topology of the network patterns. Our work serves as a starting point to explore the role of spatially correlated fluctuations and presents a route towards noise-induced phenomena in active matter.
\end{abstract}

\maketitle

Quantum fluctuations and thermal fluctuations are deeply rooted in quantum physics and in statistical mechanics. Besides opposing ordering phenomena, like magnetization, crystallization or nonequilibrium pattern formation, fluctuations can also induce fascinating phenomena at all scales, from quantum electrodynamics to living matter. 
\\In quantum electrodynamics, for instance, confined quantum fluctuations of the electromagnetic field between parallel macroscopic conducting and uncharged plates induce attractive forces between the plates (Casimir effect)\cite{Casimir-1948}. Variants of the Casimir effect arise as fluctuation-induced repulsive Casimir-Lifschitz forces (inverse Casimir-effect) \cite{Dzyaloshinskii-1961}, Casimir-Polder forces \cite{Casimir-PhysRev-1948} between neutral atoms and conducting macroscopic plates, and dynamic Casimir effects involving moving plates \cite{Dodonov-2005, Dodonov-2009}. Similarly, fluctuation-induced attractions also emerge when confining thermal fluctuations. For instance, when confining  complex correlated fluids, such as near-criticial binary liquids, or liquid crystals, between two parallel walls, fluctuation-induced attractions arise, whose strength is controlled by the thermal energy ($k_B T$) rather than by $\hbar$ that determines the uncertainty and hence the strength of fluctuations in the quantum case \cite{Fisher-1978,Hertlein-Nature-2008}. 
Other phenomena that are induced by thermal fluctuations include the emergence of entropic forces, such as entropic springs in polymer physics that stiffen when temperature increases. 
In colloids, in turn, thermal fluctuations determine the extension of the counterion cloud and hence the range of screened Coulomb interactions between different colloidal particles. These electrostatic repulsions typically compete with the Van der Waals attractions between the colloids that originate from fluctuations in the electron distribution and the corresponding dipole moments of all the atoms that make up the colloidal particles. Thus, in some sense, the stability of colloidal suspensions depends on the competition between thermal fluctuations and quantum fluctuations.
Finally, in Brownian ratchets, the interplay of thermal fluctuations and external driving forces induces a directed particle current that typically vanishes at zero temperature \cite{Haenggi-RewModPhys-2009,Reimann-PhysRep-2002, Reichhardt-AnRevCondMatPhys-2017}.\\

\begin{figure*}
 \centering
 \includegraphics[height=3.5cm]{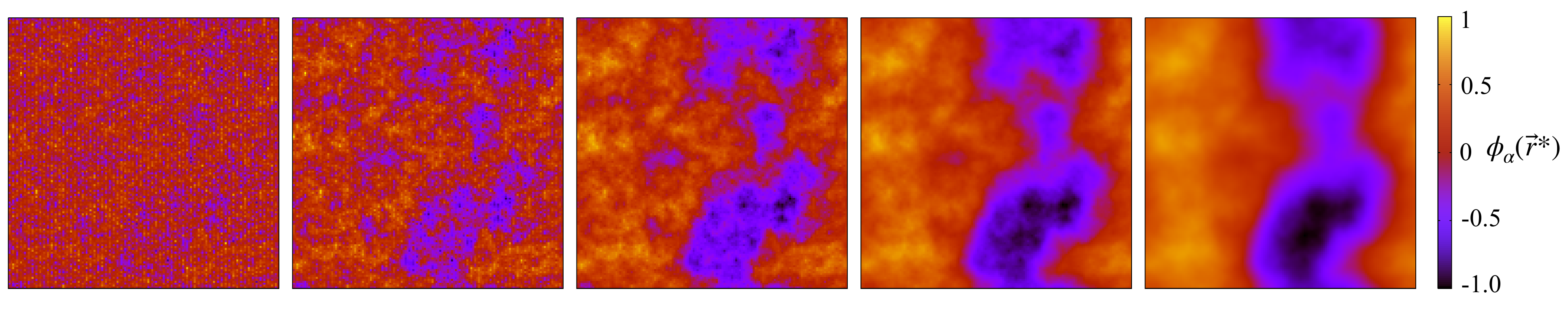}
 \caption{Exemplary realizations of the random field $\phi_{\alpha}(\vec r^*)$ for different values of $\alpha$ (from left to right: $1.0,2.0,3.0,4.0$ and $5.0$) in a simulation box of size 
 $140\times140$ in units of the particle size $\sigma$ (see SI for details).}
 \label{rf}
\end{figure*}

In some cases, interesting phenomena can also occur due to athermal fluctuations. Such fluctuations occur for instance in active matter systems, comprising self-driven units, such as self-propelled particles. Here, following the Active Ornstein Uhlenbeck model \cite{Bonilla-PhysRevE-2019,Martin-PhysRevE-2021,Nguyen-JPhysCondMat-2022,Sprenger-JPhysCondMat-2023}, activity can be understood as temporally correlated noise. It has been shown that activity, or colored noise, can induce ordering phenomena 
such as motility-induced phase separation. Similarly, in the presence of additional short-ranged attractions, active particles can form dynamic or living clusters that dynamically reform and break apart at low density \cite{Palacci-Science-2013,Theurkauff-PRL-2012,Buttinoni-PRL-2013,Ginot-NatComm-2018}, for which the interplay of activity (or colored noise), thermal fluctuations and short range attractions are likely to play a key role \cite{Palacci-Science-2013, Mognetti-PRL-2013,Liebchen-JChemPhys-2019,Liebchen-JPhysCondMat-2022}.
Other remarkable fluctuation-induced phenomena occur when an active system serves as a bath for larger (passive) objects. In this case, it has been observed that long-range forces can emerge between the objects (or between parallel plates) if the active system is chiral \cite{Grober-Nature-2023,Torrik-PRE-2021,Batton-SoftMatter-2024,Li-NewJPhys-2023}, leading to odd hydrodynamics. 

While temporal noise correlations have been widely studied in the literature \cite{Fodor-PRL-2016, Farage-PRE-2015, Paoluzzi-PRL-2024}, in the present work, we focus on spatial correlations \cite{Frechette-Arxiv-2024, Zheng-PRE-20242} and ask: What is the influence of spatial correlations in the fluctuations on the collective behavior of active systems? Such correlations could occur, for instance, when active particles with an intrinsic magnetic \cite{Urenna-PRL-2023,Novak-JPhysCondMat-2015,Han-LabChip-2021, Vidal-Urquiza-PRE-2017,Baraban-Nanoscale-2013} or electric dipole moment \cite{Liao-SoftMatter-2020,Kogler-EPL-2015} are subject to a fluctuating external magnetic or electric field that affects the orientations of adjacent particles in the system in the same way. Likewise for asymmetric Quincke-rollers \cite{Quinke-1896} an external electric field induces (chiral) active motion \cite{Imamura-AdvTheoSim-2023,Zhang-NatComm-2020, Garza-SciAdv-2023,Jones-1984,Sahu-JCIS-2024}; if the external field fluctuates, the particles experience spatially correlated fluctuations since their speed (and self-rotation frequency) depend on the external field-strength. Also when active colloids are exposed to a light intensity field ("motility field" for phototactic colloids \cite{Lozano-NatComm-2016,Jahanshahi-CommPhys-2020,Lozano-SoftMat-2019}) that varies irregularly in space but features a characteristic correlation length, they experience effective forces and torques that depend on their spatial location and may in some cases be approximated as correlated noise. 

As a starting point to explore the influence of spatially correlated noise on the collective dynamics of active particles, we introduce a minimal model that describes linear and chiral self-propelled particles. Based on systematic numerical simulations with different correlation lengths for the fluctuations, we find that spatial noise correlations supports flocking, as opposed to uncorrelated noise. Beyond that, perhaps surprisingly, we find that for chiral active particles, if fluctuations are correlated over sufficiently long distances, a new pattern arises that takes the form of a percolated network. Here, the particles are locally strongly aligned, but not globally, and coherently move in circles while forming globally connected percolating structures. We emphasize that this pattern is fluctuation induced and disappears both for uncorrelated noise and in the complete absence of fluctuations. 

\section{Model}

\begin{figure*}[!t]
 \centering
 \includegraphics[height=17.5cm]{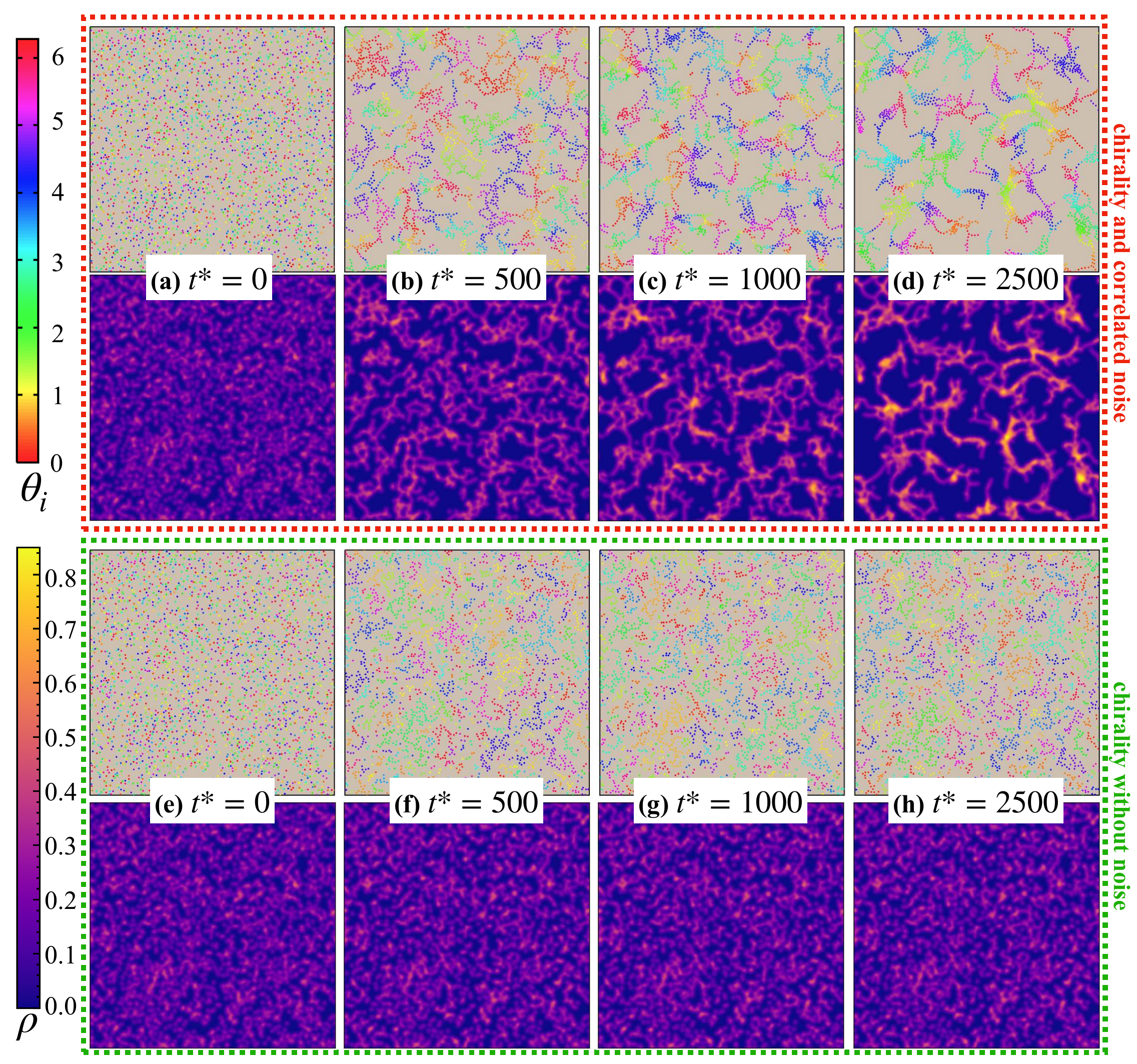}
 \caption{Noise induced network patterns: Simulation snapshots and corresponding density fields created with AMEP \cite{AMEP} for $N=2500$ particles and a correlated random field with $\alpha=4.0$. The colorbars indicates the particle orientations and the local density $\rho$ (number of particles per unit volume), respectively. (a)-(d) For $\omega^*=3.0$ and $a^*=1.0$ the particles self-organize into a percolated network pattern. In contrast to the system with spatially correlated noise, for $\omega^*=3.0$ and $a^*=0.0$ the particles do not form any structures or networks (e)-(h). Other parameters are: $\Phi_0=0.1$, $\kappa^*=1.0$, $\epsilon^*=1.0$.}
 \label{snap_01}
\end{figure*}

To explore the role of spatially correlated noise on the collective behavior of active particles, we consider a smooth variant of the Vicsek model, describing $N$ aligning active Brownian particles with a self-propulsion speed $v_0$. Optionally, for later reference, we allow the particles to be chiral ($\omega\neq 0$).
Following refs.~\cite{Liebchen-PRL-2017, Levis-JPhysCondMat-2018, Kruk-PRE-2018,Kruk-PRE-2020,Negi-PRR-2023,Caprini-CommPhys-2024}, we 
describe the dynamics of the $i$-th particle ($i=1,...,N$) with the following overdamped Langevin equations in two spatial dimensions. These equations comprise two types of fluctuations. First, 
ordinary Brownian noise due to a thermal bath (implicit solvent), which is represented by the Brownian translational and rotational diffusion coefficient $D$ and $D_r$, respectively, together with the Gaussian white noise variables $\vec \eta_i$ and $\xi_i(t)$ which have zero mean and unit variance. Second, spatially correlated noise due to a fluctuating external field, which is represented by the term $a \phi_{\alpha}(\vec r_i,t)$, where $\phi_{\alpha} \in [-1,1]$ and $a$ characterizes the strength of this field. The dimensionless parameter $\alpha$ characterizes the spatial correlation length of the fluctuating field (see Fig. \ref{rf} and SI for details on the construction of $\phi_{\alpha}$). For simplicity, we assume that the external field only acts on the orientational degrees of freedom of the particles (see SI for an exemplary exploration of correlated fluctuations acting on the center of mass coordinates of the particles).

\begin{align}
 \dot{\vec r}_i &= v_0\hat n(\theta_i) + \frac{1}{\gamma}\vec F_i + \sqrt{2D} \vec \eta_i \label{eq:1}\\
 \dot{\theta_i} &= \omega + \frac{\kappa}{\pi R_{\theta}^2} \sum_{j\in \partial_i} \sin(\theta_j - \theta_i) + a\phi_{\alpha}(\vec r_i,t) + \sqrt{2D_r} \xi_i \label{eq:2}
\end{align}

Here $\hat n(\theta_i)=(\cos\theta_i, \sin\theta_i)$ is the self-propulsion 
direction of the $i$-th particle and 
$\vec F_i = -\sum_{\genfrac{}{}{0pt}{} {j=1}{j \neq i}}^{N} \nabla_{\vec r_i} U_S(r_{ij})$
is the steric interaction force (volume exclusion)  
acting on the $i$-th particle, which derives from the interaction potential $U_S(r)= \epsilon \left(\frac{\sigma}{r}\right)^{12}$ for $r \leq 2^{1/6}\sigma$ and $U_S(r)=0$ else, where $\sigma$ is the (soft) particle diameter, 
$\epsilon$ is the interaction strength, $\gamma$ is the Stokes drag coefficient, and
$r_{ij}=|\vec r_i - \vec r_j|$.
Further, in Eq. (2) $\omega$ represents the bias in the orientation of the particles, i.e., $\omega=0$ for straight/linearly moving active Brownian particles and $\omega\neq 0$ for chiral active particles. The second term in Eq. (2) represents alignment interactions of strength $\kappa$, with the sum running over all particles within a disk of radius $R_{\theta}=2\sigma$ around the $i$-th particle.
 
For vanishing correlation length, the $\phi_{\alpha}$-field reduces to Brownian fluctuations in the $\theta$-variable. For simplicity, in the following we neglect  additional thermal fluctuations, which are rather unimportant in most of the cases which we explore in the following (see SI). 

\begin{figure*}[t!]
 \centering
  \includegraphics[width=0.90\textwidth]{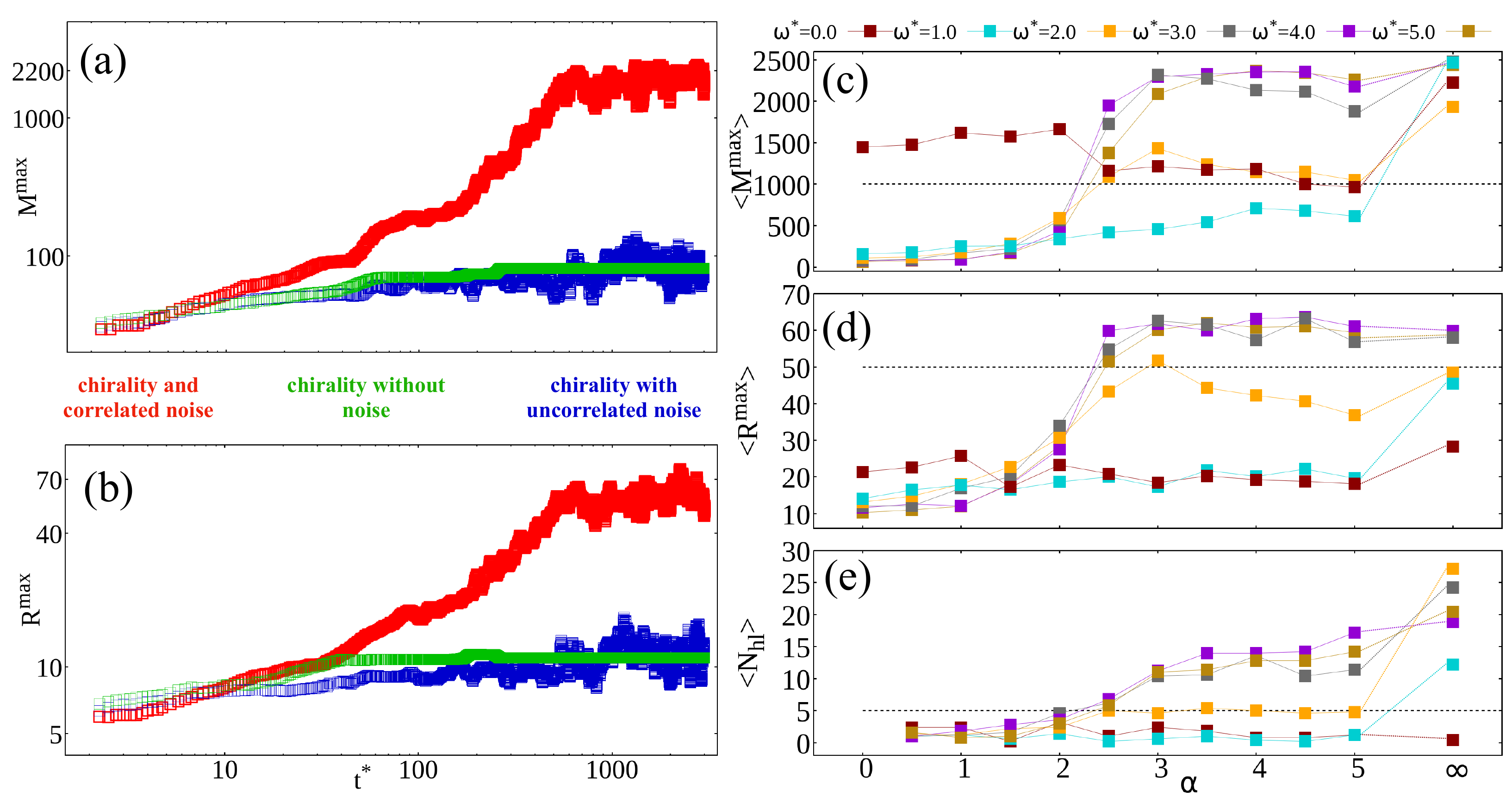}
 \caption{Time evolution of (a) the size of the largest connected structure in terms of the number of contained particles $M^{\mathrm{max}}$ and (b) its radius of gyration $R^{\mathrm{max}}$. The red lines show the case, where a correlated random field acts on the particle orientations $(\omega^*=3.0, \alpha=4.0, a^*=1.0)$ and the blue ones with uncorrelated noise $(\omega^*=3.0, \alpha=0.0)$, while the green lines correspond to the deterministic system without any fluctuations $a^*=0.0$ also with $\omega^*=3.0$. Other parameters are: $\Phi_0=0.1$, $\kappa=1.0$, $\epsilon^*=1.0$. (c) Averaged largest cluster size $\langle M^{\mathrm{max}}\rangle$, (d) corresponding radius of gyration $\langle R^{\mathrm{max}}\rangle$ and  average number of holes $\langle N_{\mathrm{hl}}\rangle$ as a function of the noise correlation exponent $\alpha$ for different values of $\omega^*$ as shown in the legend. The black dotted lines indicate the threshold defined in the section ``Non-equilibrium state diagram''.}
\label{sizes}
\end{figure*}

Now, expressing lengths and times in units of $\sigma$ and $\sigma/v_0$, respectively, the equations of motion reduce to  $\dot{\vec r}_i^* = \hat n_i(t^*) -\nabla_{r^*} U^*(r^*)$ and  
$\dot{\theta}_i^* = \omega^* + \kappa^*\sum_{j\in\partial_i}\sin(\theta_j^* - \theta_i^*) + a^*\phi_{\alpha}(\vec r_i^*,t^*)$, where $U_S^*(r^*) = \epsilon^* \left(\frac{1}{r^*}\right)^{12}$. The six remaining dimensionless control parameters are: the effective alignment strength $\kappa^*=\frac{\kappa\sigma}{\pi R_{\theta}^2v_0}$, the strength of the steric repulsions $\epsilon^* = \frac{\epsilon}{\gamma v_0 \sigma}$, the packing fraction $\Phi_0 = \frac{N\pi\sigma^2}{4L^2}$ where $L$ is the length of the quadratic simulation box which features periodic boundary conditions, the reduced rotation rate 
$\omega^* = \frac{\omega\sigma}{v_0}$ and two parameters characterizing the spatially correlated noise. These are 
the reduced amplitude of the random field $a^*=\frac{a\sigma}{v_0}$ and the dimensionless parameter $\alpha$. 
We now solve the equations of motion with a forward Euler method (Euler-Maruyama scheme for handling the fluctuations) for $N=2500$ particles with random initial positions, but without any overlap, and orientations (see Fig. \ref{snap_01}(a) and (e)). Here, we fix the packing fraction to $\Phi_0=0.1$, the interaction parameters $\epsilon^*=1.0$ and $\kappa^*=1.0$ (flocking regime for $\omega^*=0$) and the amplitude of the random field to $a^*=1.0$ (unless stated differently) and systematically vary the parameters $\omega^* \in [0,5]$ and $\alpha \in [0,5]$.

\section{Fluctuation induced network patterns}

What is the role of correlated noise on the large scale collective behavior of the particles? Does it only oppose well-known structures such as flocking for non-chiral particles and pattern formation for chiral particles, or can it also induce new structures? Let us now explore these questions step by step.

First, for $\omega^*=0$ and $\alpha=4.0$ (see Fig. S1(a)-(d) and Movie01 in the SI and Fig. \ref{phasediagram} bottom row) the particles align with their neighbors and aggregate in clusters (flocks). This behavior closely resembles well-known behavior in the Vicsek model with uncorrelated noise at low-to-moderate densities \cite{Kruk-PRE-2020,Vicsek-PRL-1995,Chate-EurPhysJ-2008,Toner-PRL-1995,Bertin-PRE-2006,Ihle-PRE-2011, Gregoire-PRL-2004,Solon-PRL-2015} suggesting that spatial noise-correlations do not have too much influence on the collective behavior of the system. 

Second, for aligning chiral active particles with a moderate rotation frequency ($\omega^*=1.0$), it is well known that in the case of uncorrelated noise rotating clusters featuring internal alignment emerge at sufficiently high density, that have been called rotating microflocks \cite{Liebchen-PRL-2017, Levis-JPhysCondMat-2018} (or polar rotating packets in ref. \cite{Ventejou-PRL-2021}). At such intermediate frequencies, $\omega^*=1.0$, the presence of noise correlations does not have much influence on the emerging structures, apart from enhancing their size (see Fig. S1(e)-(h) and Movie02 in the SI). 
 
In stark contrast, for larger chirality, $\omega^*=3.0$, spatially correlated noise induces a completely different type of pattern. We observe, that elongated structures emerge at early times that connect with each other (see Fig. \ref{snap_01}(a)-(d) and Movie03 in the SI). In fact, while for uncorrelated noise it is well known that the size of rotating microflocks $l$ decreases when $\omega$ increases, as $l \propto 1/\omega \propto 1/\omega^*$ (see also data points for $\alpha=0$ in Fig. \ref{sizes}(c) and (d)), for correlated noise, the size of the connected structures strongly increases with $\omega^*$. That is, unlike for microflock patterns that emerge for uncorrelated noise and that feature a characteristic length scale that is independent of the system size, spatially correlated noise induces a qualitatively different pattern that spans the entire system (i.e. it scales with the system size). Within the networks, particles are locally aligned (see colors in Fig. \ref{snap_01}(a)-(d)) and move only locally on small circles.

\begin{figure*}[!t]
 \centering
 \includegraphics[height=8.0cm]{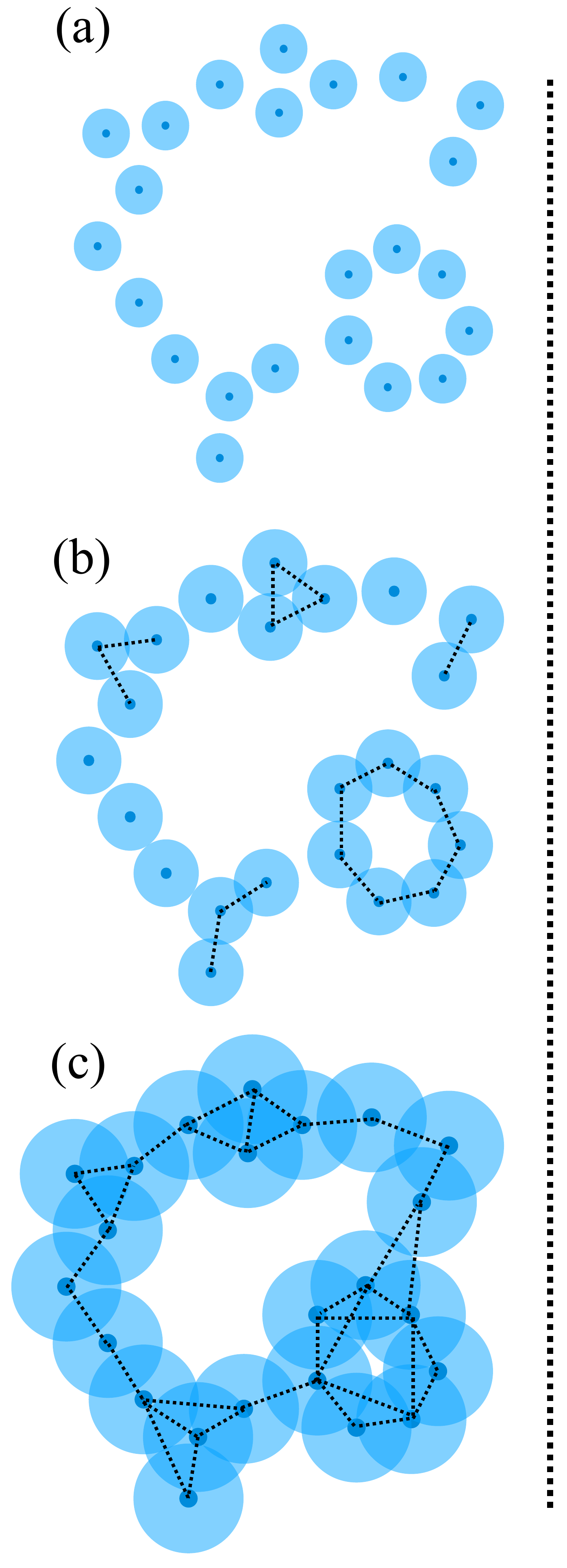}\includegraphics[height=8.0cm]{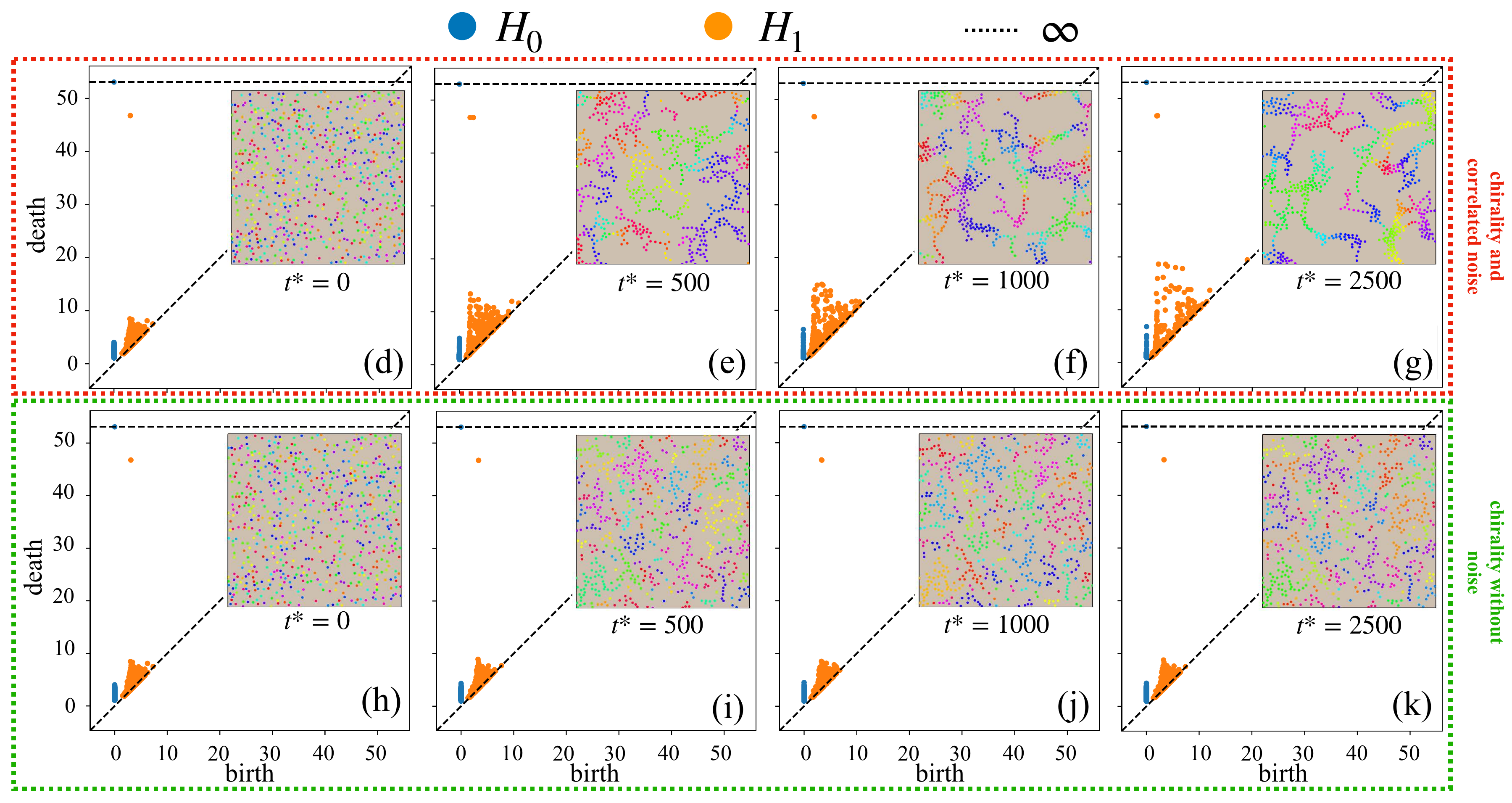}
 \caption{Topological data analysis: 23 data points form different Vietoris-Rips complexes for three different proximity parameters $d$, shown by the lightblue disk around each point. (a) For small $d$, their is no overlap and
therefore no connections between the points. Increasing the disk diameter $d$ forms new ``connections'' and holes and finally (c) all particles build one connected component. Persistence diagrams for $\omega^*=3.0$ and $\alpha=4.0$ ((d)-(g)) and for the case of a vanishing random field $\phi_{\alpha}(\vec r^*,t)$ ((h)-(k)) corresponding to the simulation snapshots shown in Fig. \ref{snap_01} (a)-(d) and (e)-(h), respectively. The insets show more detailed snapshots. Note, that in all diagrams, the two points for high death values are artifacts of the periodic boundary conditions and do not have a physical meaning.}
\label{pd_diagrams}
\end{figure*}

To test if the network-patterns are fluctuation-induced, or if they also occur in the complete absence of fluctuations, we perform additional simulations without any noise, i.e. for $a^*=0$, but with the same initial conditions (see Fig. \ref{snap_01}, (e)-(h) for snapshots and Movie04 in the SI). In this case, the particles also locally align with each other, but at least on the timescale of our simulations, there is no signature of the formation of any large clusters or even system-spanning networks. 

To quantify this, we now calculate the size of the largest cluster $M^{\mathrm{max}}$, in terms of the number of contained particles, and the corresponding radius of gyration $R^{\mathrm{max}}$ using OVITO \cite{ovito}. We choose a threshold distance of $r_c^*=2.0$ such that particles with a distance smaller than $r_c$ belong to the same cluster. While in the presence of correlated noise ($a^*=1$) we have $M^{max}\sim 2000$ and $R^{max}\sim 55.0\sigma$ (red lines in Fig. \ref{sizes}(a) and (b)), i.e. the largest cluster is comparable to the system size, in the absence of fluctuations ($a^*=0$) the largest cluster is much smaller (green line in Fig. \ref{sizes}(a) and (b)). Overall, our observations suggest that the observed network patterns neither emerge for uncorrelated noise nor in the complete absence of noise, but they require spatially correlated noise to emerge.

\section{Characterizing network patterns by topological data analysis}

A characteristic feature of the network patterns is that they comprise holes, i.e. void spaces that are surrounded by layers of particles. In the following, we first directly compute the average number of holes $\langle N_{hl}\rangle$ using the same threshold-distance as for our cluster-size-analysis ($r_c^*=2.0$, see SI for details) and then proceed with a systematic topological data analysis, which does not require choosing a threshold-distance.

Following the first approach and calculating $\langle N_{hl} \rangle$ at the end of our simulations yields Fig. \ref{sizes}(e). Here, we observe that for weakly correlated fluctuations, $\alpha < 2.0$, the number of holes is small for all simulated $\omega^*$ values. Increasing $\alpha$ leads to a larger number of holes, in particular for $\omega^* \geq 3$. This coincides with our qualitative observations (Fig. \ref{snap_01}). While the number of holes as calculated so far sensitively depends on the chosen distance threshold $r_c$, we now proceed with a systematic characterization of the network patterns based on a topological data analysis (TDA) and use the persistent homology method. These tools have their origin in algebraic topology and give us the opportunity to get deeper insights in the topological structure of the network patterns. First, we give a short and pragmatic introduction on selected features of TDA \cite{Topaz-PLOS-2015, Chazal-2021,Bhaskar-SoftMatter-2021} (see also \cite{Edelsbrunner-ConMath-2008,Carlsson-AMS-2009} for mathematical details) before analyzing our data.  

The homology of data (in our case the positions of particles) can be characterized by creating connections between those data points (particle positions) that have a distance of at most $d$, where $d$ is called proximity parameter. To visualize this, one can imagine to draw an n-dimensional sphere (disk in two dimensions) of diameter $d$ around each data point (particle position) and connect those data points for which the emerging spheres overlap (Fig. \ref{pd_diagrams}(a)-(c)). One then systematically varies $d$ and explores the topological structures (Vietoris-Rips complexes) that emerge (``birth'') and the ones that disappear (``death'') for each $d$ (see Fig. \ref{pd_diagrams}). In particular, a persistence diagram (PD) can then be obtained by varying the proximity parameter $d$ and tracking the number of connected components $H_0$ ($0$-dimensional topological holes, e.g. $H_0=23$ in Fig. \ref{pd_diagrams}(a), $H_0=10$ in Fig. \ref{pd_diagrams}(b) and $H_0=1$ in Fig. \ref{pd_diagrams}(c)), topological loops $H_1$ ($1$-dimensional topological holes , also called $1$-simplexes, e.g. $H_1=0$ in Fig. \ref{pd_diagrams}(a) and $H_1=2$ in Fig. \ref{pd_diagrams}(b)) or more generally the homology in the resulting Vietoris-Rips complex. A PD then displays the ``birth'' of a topological structure on the horizontal-axis and its ``death'' on the vertical-axis of a two dimensional coordinate system (Fig. \ref{pd_diagrams}). For $H_0$, the birth value is always zero, since every single point corresponds to one connected component. Accordingly, the corresponding death value (blue symbols in Fig. \ref{pd_diagrams}(d)-(k)) tells us, for which $d$ two (or more) connected components (isolated or connected particles) form one larger complex. Topological holes $H_1$ emerge for $d$ values, where connected points form a ring and the corresponding death value mimics exactly that $d$ for which this hole is completely filled out and therefore gives us information about the size of this hole.

While in general any higher-dimensional topological holes can be considered (with an increasingly expensive computational effort), we focus in this work on $0$-dimensional and $1$-dimensional holes (Fig. \ref{pd_diagrams}), which allow for a useful interpretation.

\subsection*{Hole size analysis based on Vietoris-Rips complexes}

\begin{figure}[h!]
\centering
 \includegraphics[height=5.2cm]{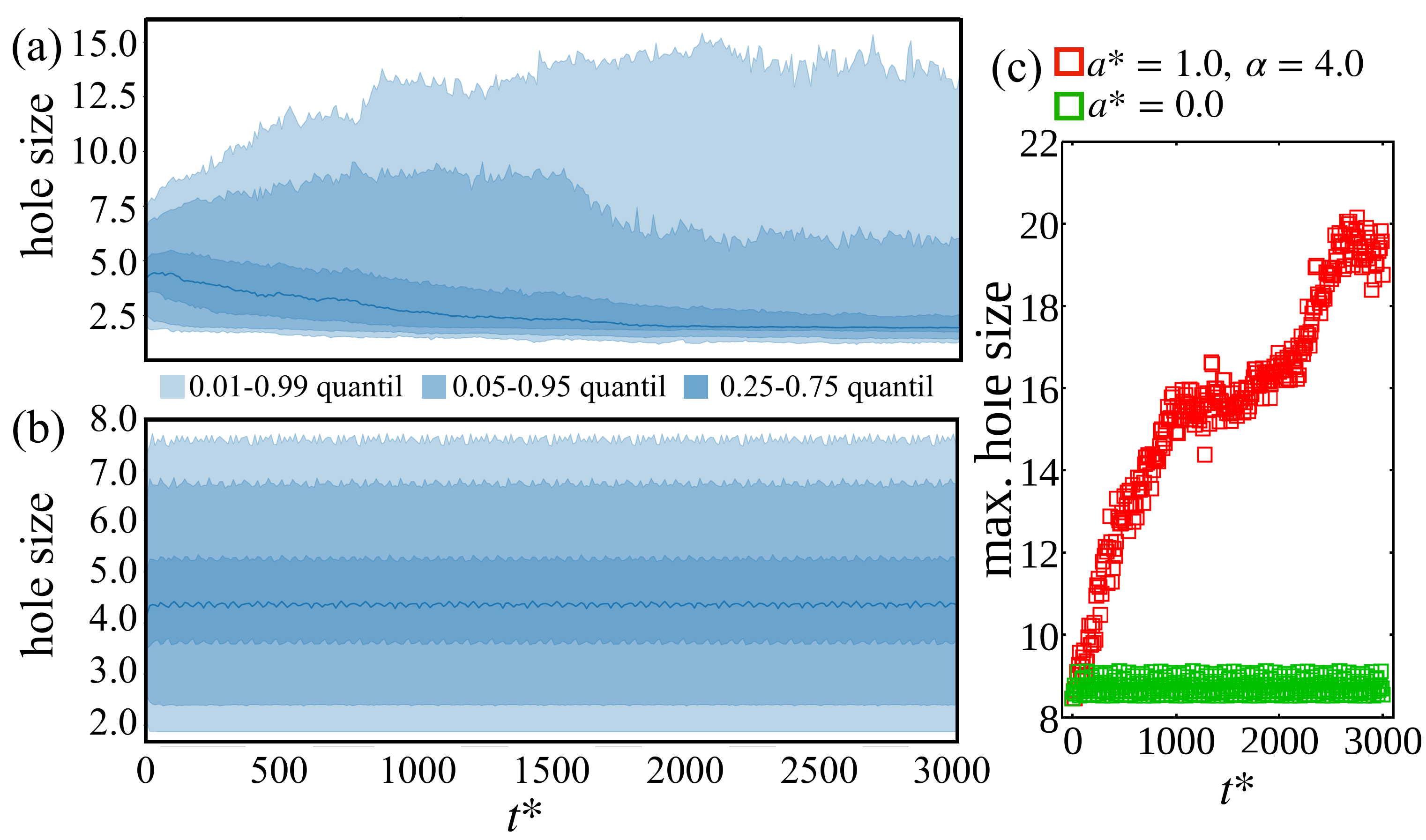}
 \caption{Hole size distribution extracted from the death values in the PDs belonging to birth values $<5$. The shades of blue show different quantils of the distribution as indicated in the key. (a) $\omega^*=3.0$ and $\alpha=4.0$. (b) $\omega^*=3.0$ and $a^*=0.0$. Panel (c) shows the largest hole size as a function of time $t^*$ for parameters shown in the key; $\omega^*=3.0$ for both data.}
\label{hole_size}
\end{figure}

We now analyze our simulation data (using the code provided in\cite{scikittda2019}), based on the PDs (Fig. \ref{pd_diagrams}) and first have a look at holes given by $H_1$. For $\omega^*=3.0$ and $\alpha=4.0$ (Fig. \ref{pd_diagrams}(d)-(g)) as time increases (left to right), we identify the emergence of holes with an increasing size, since the points in the PD show increasing death-values. We observe holes which are born for $d\lesssim 10$ and which persist up to death values of around $20$. In contrast, for $a^*=0$ (Fig. \ref{pd_diagrams}(h)-(k)), most of the $H_1$ points are close to the diagonal ``birth=death'' which represents the fact, that the system just shows holes much smaller than in the case of a correlated random field. To quantify this, we extract the hole size distribution from the PDs and plot it as a function of time (see Fig. \ref{hole_size}(a) and (b)). To do so, we take all $H_1$ points in the PD which have a birth value $<5$ and build a distribution of their corresponding death values, which essentially represents the size of a (persistent) hole. While the mean hole size is similar with ($\alpha=4.0$) and without ($a^*=0.0$) correlated noise, the system with correlated noise shows a significantly broader distribution and therefore also larger holes, which coincides with our qualitative observations and our calculation of $N_{hl}$ before. This becomes even clearer when only looking at the size of the largest hole (see Fig. \ref{hole_size}(c)). While this value is roughly constant at values $\sim 8.5$ for the system without correlated noise (green symbols), the hole size increases up to $\sim 20$ for $\omega^*=3.0$ and $\alpha=4.0$ (red symbols), which again reflects the observation of network patterns.

\subsection*{Characterizing percolation with Betti numbers}

\begin{figure}[!h]
\centering
 \includegraphics[height=3.5cm]{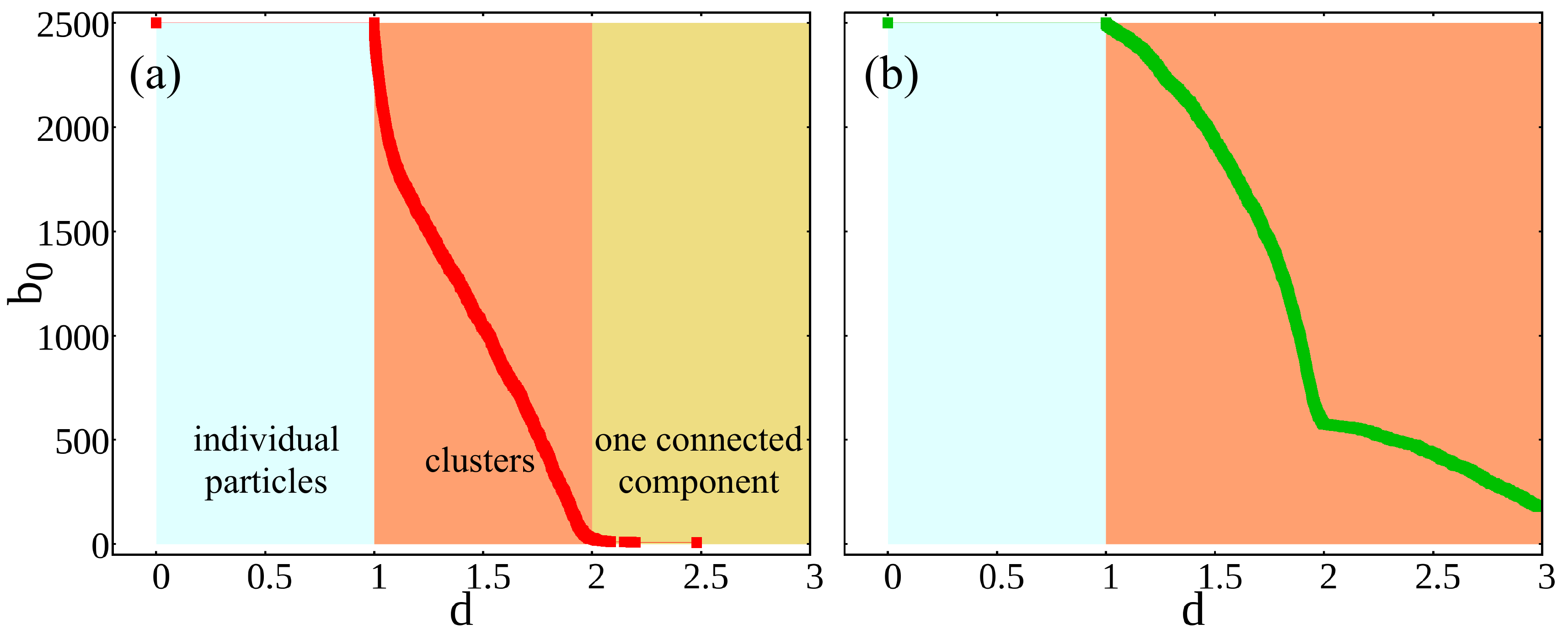}
 \caption{Betti number $b_0$ as a function of the proximity parameter $d$. (a) $\omega^*=3.0$ and $\alpha=4.0$. (b) $\omega^*=3.0$ and $a^* = 0.0$ corresponding to a deterministic system with $\phi_{\alpha}(\vec r^*,t)=0$.}
\label{betti}
\end{figure}

Further, from the PDs, we extract the so-called Betti number $b_0$. It can be read off from the persistence diagram for $0$-dimensional holes by counting the number of $0$-dimensional holes with death before $d$ on the $y$-axis. It is essentially the number of connected components of a data set, depending on the proximity parameter $d$. Fig. \ref{betti} shows Betti plots, i.e., the Betti number $b_0$ depending on the proximity parameter $d$. First we note, that for $d<1$ (disks smaller than the particles) $b_0$ just measures the number of particles in the system $(N=2500)$ (e.g. Fig. \ref{pd_diagrams}(a), lightblue area), which is reasonable, since particles do not show any significant overlap. In the presence of correlated noise (Fig. \ref{betti}(a)), for $d\gtrsim 1$ and $d\lesssim 2$, $b_0$ decreases to values close to one, which means that for $d\gtrsim 2$ more or less all particles are connected. This situation, where all particles build one connected component, arises for a vanishing random field (see Fig. \ref{betti}(b)) just for $d\gtrsim 3.5$, indicating that most particles are at a comparatively large distance to each other. This also matches with our results for the largest cluster size (Fig. \ref{sizes}(c) and (d), and shows that the previously used ad hoc choice of a critical distance of $r_c^*=2.0$ is sensible. 

\begin{figure*}[t!]
 \centering
 \includegraphics[width=0.92\textwidth]{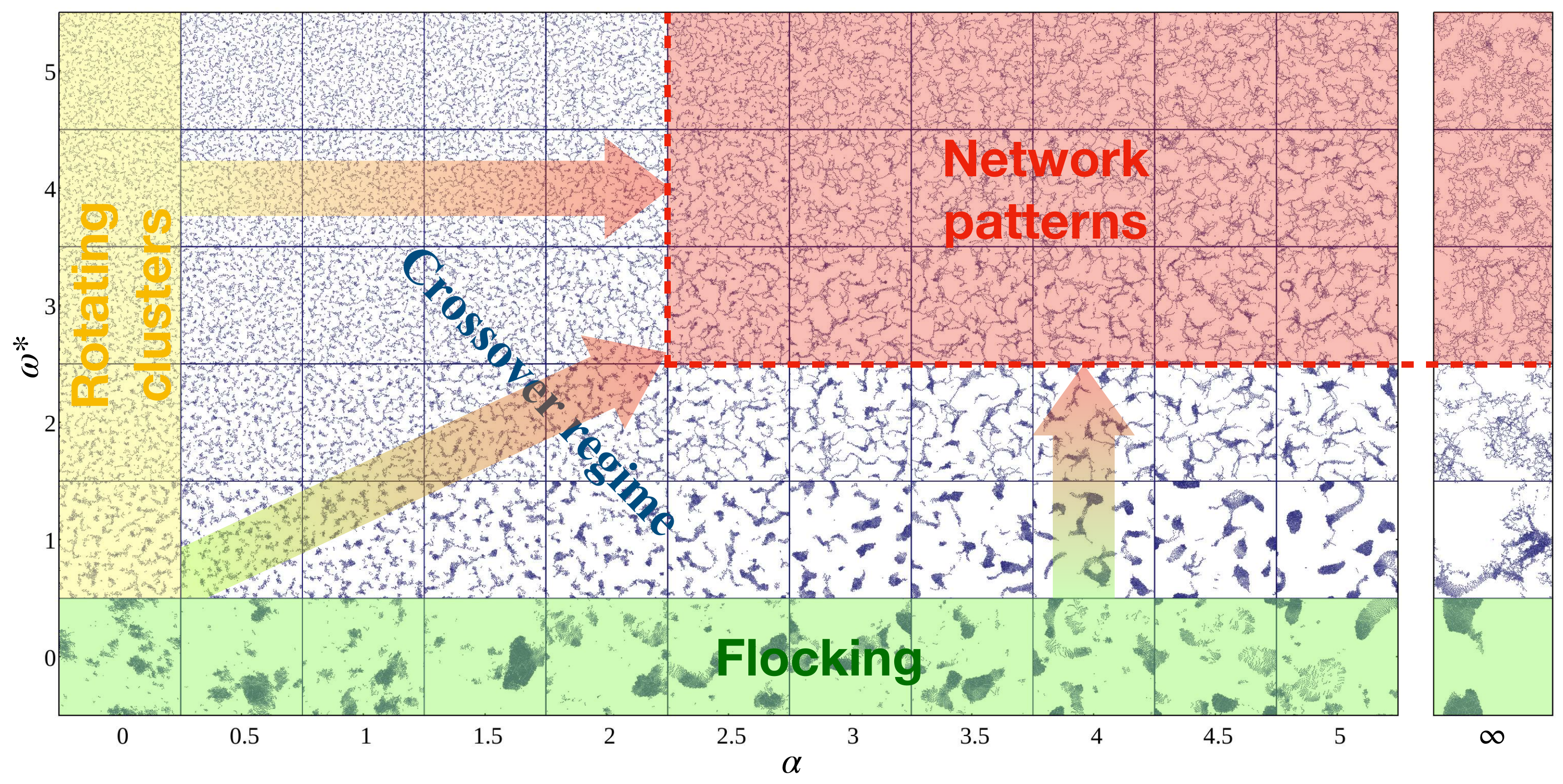}
 \caption{Non-equilibrium state diagram. The red framed area in the top right marks the regime, where we find network patterns which fulfill the criteria defined in the main text.}
\label{phasediagram}
\end{figure*}

\section{Non-equilibrium state diagram}

We now summarize our observations in a non-equilibrium state diagram (Fig. \ref{phasediagram}) in the $\omega^*$-$\alpha$-parameter plane. Here, structures in the top right (separated by the red line) comprise a connected structure of size $\langle M^{max}\rangle >10^3$ (Fig. \ref{sizes}(c)), a radius of gyration $\langle R^{max}\rangle >50\sigma$ (Fig. \ref{sizes}(d)), and has more than 5 holes (Fig. \ref{sizes}(e)). This diagram reflects that network patterns emerge only if the correlation length of the noise ($\alpha$) and the rotation frequency ($\omega^*$) is sufficiently large.

\section{Physical mechanism}

\begin{figure}[h!]
 \centering
 \includegraphics[width=0.46\textwidth]{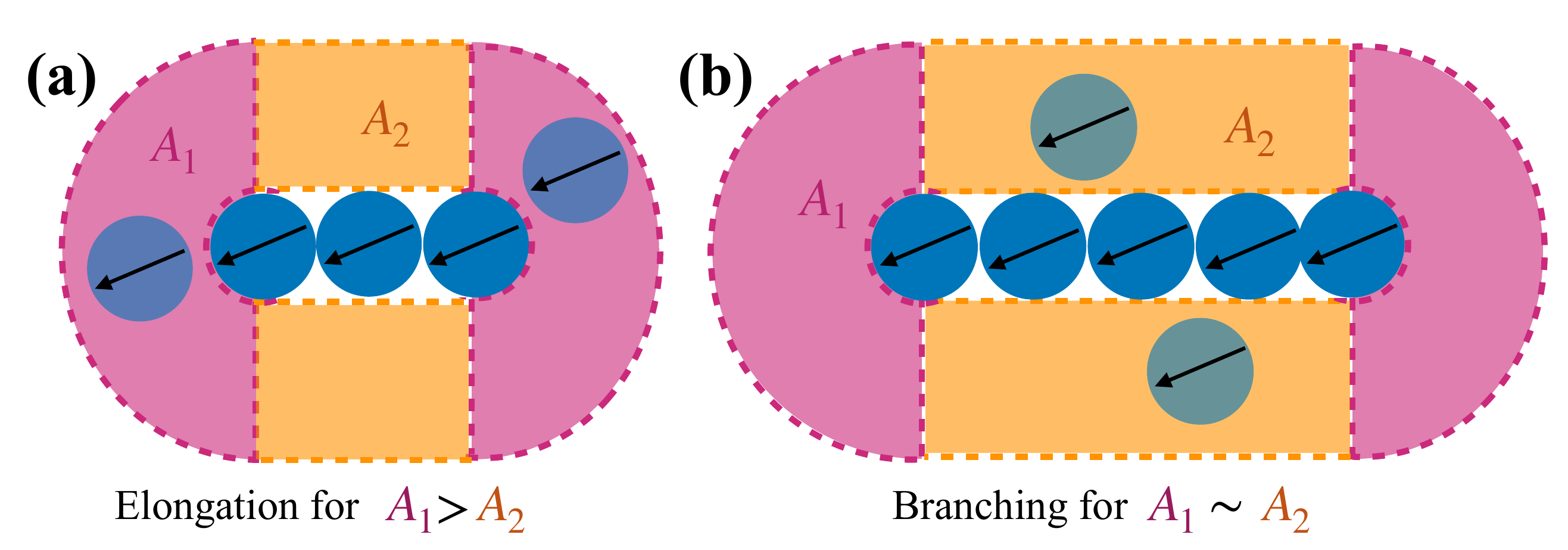}
 \caption{Illustration of the mechanism leading to the formation of elongated structures. The areas $A_1$ and $A_2$ illustrate the range for alignment interactions. Assuming a uniform distribution of neighboring particles, in (a) attachment of a particle in region $A_1$ is more likely, making the linear chain longer. Once the chain is long enough that $A_1=A_2$, branching events become equally likely.}
\label{sketch}
\end{figure}

We first discuss the question, why noise is necessary at all for the formation of network patterns. Without the random field  in Eq. (\ref{eq:2}) (and neglecting thermal noise), the system is completely deterministic, and once particles which are initially in close vicinity to each other align, the system reaches a quasistationary state, featuring a few small clusters that do not grow in time. Then, particles essentially move in a coordinated and almost time-periodic way, following small circles while keeping a certain distance to each other. In this case, the morphology of the pattern doesn't evolve in time. Second, we rule out a significant impact of the volume exclusion interactions, since even with $\epsilon^*=0$ particles form similar network patterns (see SI for exemplary snapshots). Finally, we give an argument, why particles form elongated, string-like structures in the presence of correlated noise, which then connect to system spanning networks. To see this, first we note, that spatially correlated noise doesn't oppose local alignment in contrast to uncorrelated noise. Once a dimer of two particles has formed, it therefore persists over a long time. In addition, for geometrical reasons, it is more likely for a third particle to attach at the head or the tail of the dimer then in the middle, thus leading to elongated structures (see Fig. \ref{snap_01}). Assuming a uniform density of the surrounding particles, the corresponding probabilities are given by the colored areas in Fig. \ref{sketch}, where one clearly sees that $A_1>A_2$ for short chains. Beyond a critical chain length, defined by $A_1=A_2$, branching becomes more likely than linearly extending the chain and a network pattern begins to form. Following this geometric consideration, the typical length scale of the linear chains is determined by $A_1= A_2$, which is equivalent to $\pi (2\sigma)^2 - \pi \left( \frac{\sigma}{2}\right)^2 = 3\sigma(N-1)\sigma \ \rightarrow \ N \approx 5 $, so $l_{\mathrm{c}} \gtrsim 5$ (the particle diameter $\sigma = 1$). This rough estimation is consistent with our observations in Fig. \ref{snap_01} and \ref{hole_size}. We further examine three different characteristic timescales: the lifetime of a dimer in correlated noise $\tau_{\mathrm{c}}$, the lifetime of a dimer in uncorrelated noise $\tau_{\mathrm{uc}}$ and the time it takes on average for a third particle to attach to the dimer $\tau_{\mathrm{a}}$. We find that $§\tau_{\mathrm{c}}> \tau_{\mathrm{a}} \approx 162 >\tau_{\mathrm{uc}}\approx 38$ (see SI for details). This explains why dimers form and grow in the spatially correlated noise fields, but not in the presence of uncorrelated noise.

\section{Conclusions}

We have introduced a minimal model to explore the influence of spatially correlated noise in active matter. As our key result, we have observed that under the influence of spatially correlated noise, chiral active particles self-organize in a network pattern that features a system-spanning percolated structure, intrinsic holes and local alignment of the active particles that make up the network. We have characterized the emerging network patterns using methods from topological data analysis (persistence diagrams, Vietoris-Rips complexes and Betti numbers) that have so far hardly been used in active matter. The emerging network patterns arise only in the presence of spatially correlated noise and represent an example of a fluctuation-induced phenomenon in active matter physics. Overall the present work serves as a starting point to explore the role of correlated noise in active matter and invites future studies to also systematically explore, e.g. spatially and temporally correlated noise, the impact of correlated noise  on the center of mass coordinate of the particles and also for external fields that do not only provide fluctuations but also induce body forces or torques.

\subsection*{Author contributions}
Sebastian Fehlinger: numerical simulations, data analysis, visualization and writing (original draft, review and editing). Kai Cui: topological data analysis and visualization. Arooj Sajjad: numerical simulations, data analysis. Heinz Koeppl: topological data analysis, review and editing. Benno Liebchen: conceptualization, project administration and writing (original draft, review and editing). 

\subsection*{Conflicts of interest}
There are no conflicts to declare.

\subsection*{Data availability}

All algorithms used in this work are explained in the Supplementary Information. The code will be made available upon reasonable request. Data analysis involves the Active Matter Evaluation Package that is available at: https://amepproject.de

\subsection*{Acknowledgements}
We thank Kay-Robert Dormann for the useful discussions and for the help with AMEP. 







\providecommand*{\mcitethebibliography}{\thebibliography}
\csname @ifundefined\endcsname{endmcitethebibliography}
{\let\endmcitethebibliography\endthebibliography}{}


\newpage

\setcounter{section}{0}
\setcounter{equation}{0}
\setcounter{figure}{0}
\setcounter{table}{0}
\setcounter{page}{1}
\makeatletter
\renewcommand{\theequation}{S\arabic{equation}}
\renewcommand{\thefigure}{S\arabic{figure}}
\renewcommand{\bibnumfmt}[1]{[S#1]}
\renewcommand{\citenumfont}[1]{S#1}

\begin{widetext}
\begin{center}
\textbf{Supplementary Information to “Fluctuation induced patterns in active particles with spatially correlated noise”} \\
Sebastian Fehlinger, Kai Cui, Arooj Sajjad, Heinz Koeppl, and Benno Liebchen
\end{center}

\section{Correlated random fields}

We use the method described in \cite{Liu-GeoSci-2019,Beumann-PRE-2013,Carron-MNRAS-2014}. That is, we start with a Gaussian white noise field with zero mean and unit variance $\phi_{\mathrm{white}}(\vec r)$. The second step is to Fourier transform this field to get $\tilde{\phi}_{\mathrm{white}}(\vec k)$. As a third step, we choose $P(\vec k) = |\vec k|^{-\alpha}$. Finally, we define $\tilde{\phi}(\vec k) = P^{1/2}(\vec k) \tilde{\phi}_{\mathrm{white}}(\vec k)$ and do the inverse Fourier transformation to get the correlated field $\phi(\vec r)$ in real space. We note, that when defining $\tilde{\phi}(\vec k)$ in this way, $P(\vec k)$ is the Fourier transform of the correlation function $C(\vec r,\vec r') = \langle \phi(\vec r)\phi(\vec r')\rangle$, which shows, that the random field exhibits a characteristic distribution of correlation lengths, that is determined by the chosen power law exponent $\alpha$. Thus, the larger the value of $\alpha$, the stronger is the presence of longer ranged correlations in $\phi(\vec r)$. To guarantee periodic boundary conditions, we choose $\vec k=\frac{2\pi\vec r}{L}$. Numerically, we generate a lattice on our simulation box of size $L\times L$. The distance between two lattice sites is equal to the particle diameter $\sigma$ (which is sufficient in the absence of temporal correlations of $\phi$). Then, for every grid point we draw a random number from a Gaussian distribution with zero mean and unit variance (white noise). After the Fourier transformation (using the FFTW package in C \cite{FFTW}), these random numbers are multiplied with the square root of the power spectrum before using the inverse Fourier transformation to get back to real space. Different realizations of the field for different $\alpha$ are shown in Fig. 1 in the main text.

\section{Numerical simulations}

For solving the equations of motions we use a straight forward Euler-Maruyama scheme, which is reasonable, since the spatially correlated random field is uncorrelated in time, so $\langle \phi(\vec r^*,t^*) \phi(\vec r^*,t^{*\prime} \rangle) \sim \delta(t^*-t^{*\prime})$. The time discretized equations are (already in dimensionless units as in the main text)

\begin{align*}
\vec r_i^* (t^* + \Delta t^*) &= \vec r_i^* (t^*) +  \hat n_i(t^*)\Delta t^* -\Delta t^* \nabla_{r^*} U^*(r^*) \\  
\theta_i^*(t^*+\Delta t^*) &= \theta_i^*(t^*) + \omega^*\Delta t^* + \Delta t^* \kappa^*\sum_{j\in\partial_i}\sin(\theta_j^* - \theta_i^*) + \sqrt{\Delta t^*}a^*\phi_{\alpha}(\vec r_i^*,t^*).
\end{align*}

In all simulations we use a time step of $\Delta t^*=10^{-5}$.

\subsection{Flocking behavior}

As already discussed in the main text, for comparatively small rotational frequencies, the system shows flocking, similarly as in the Viscek model, see Fig. \ref{snap_si_flocks} for snapshots.

\begin{figure}
 \includegraphics[width=\textwidth]{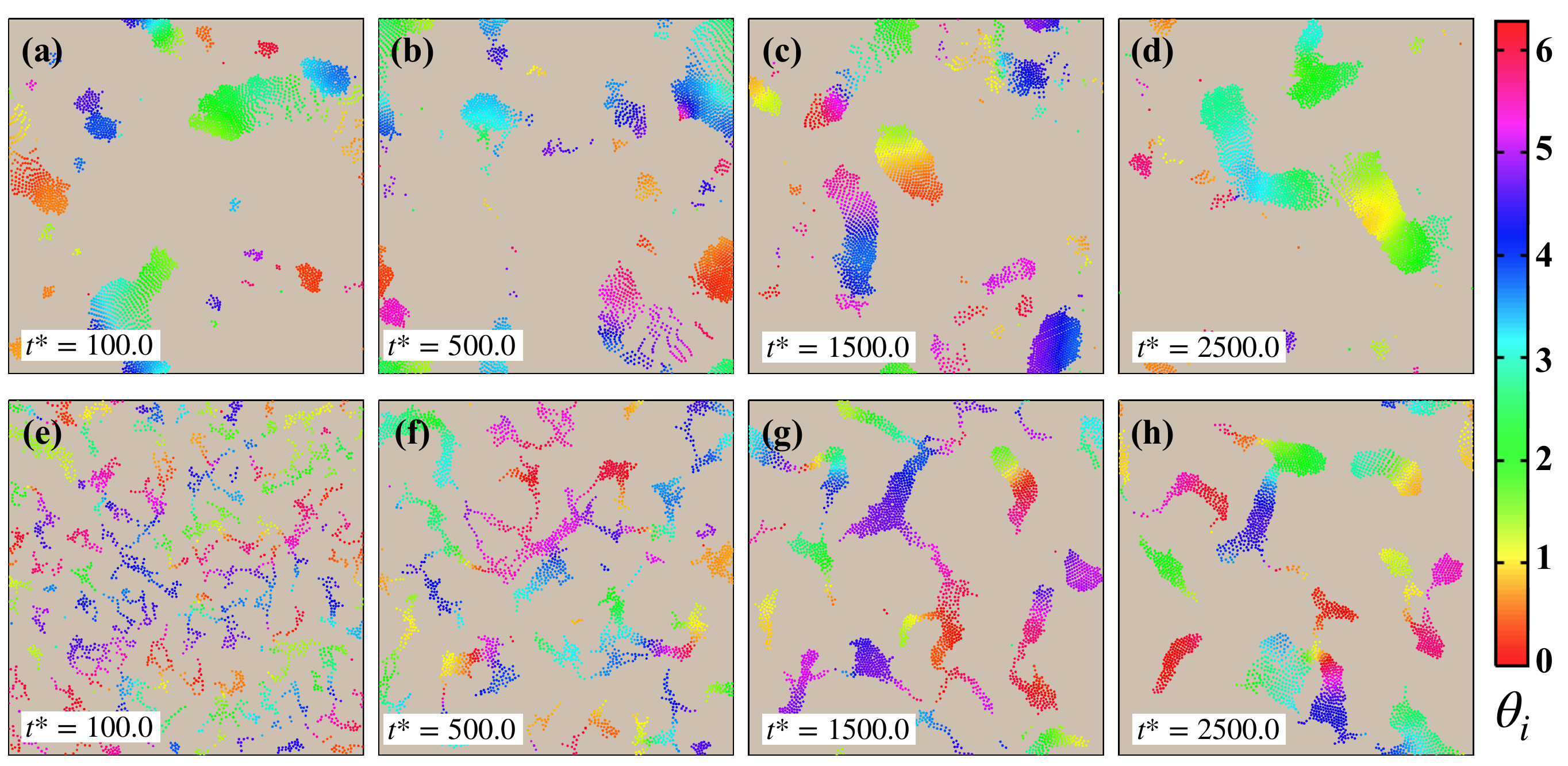}
 \caption{Snapshots for $N=2500$ particles and a correlated random field with $\alpha=4.0$. The colorbar indicates the orientation $\theta_i$ of each particle. (a)-(d) For $\omega^*=0.0$ we observe flocking as described by the Viscek model. The spatial correlations sometimes pull particles within a flock apart from each other. (e)-(h) Increasing the rotational frequency to $\omega^*=1.0$ leads to the formation of elongated structures which build again flocks for longer times. Other parameters are: $\Phi_0=0.1$, $\kappa^* =1.0$, $\epsilon^*=1.0$ and $a^*=1.0$.}\label{snap_si_flocks}
\end{figure}

\subsection{Robustness of network patterns against additional Gaussian white noise}

To test the robustness of the network patterns (Fig. 2 in the main text) against thermal fluctuations, we now include additional Gaussian white noise in the equations of motion of all particles, which corresponds to $D,D_r \neq 0$ in Eqs. (1),(2) in the main text. In dimensionless form, the equations now read:

\begin{align*}
 \dot{\vec r}_i^*(t^*) &= \hat n_i(t^*) -\nabla_{r^*} U^*(r^*) + b^*\vec\eta_i(t^*) \\
 \dot{\theta}_i^*(t^*) &= \omega^* + \kappa^*\sum_{j\in\partial_i}\sin(\theta_j^* - \theta_i^*) + a^*\phi_{\alpha}(\vec r_i^*,t^*) + c^*\xi_i(t^*),
\end{align*}

where $\vec \eta$ and $\xi$ describe white noise with zero mean and unit variance with parameters $b^*=\sqrt{\frac{2D}{\sigma v_0}}$ and $c^*=\sqrt{\frac{2D_r\sigma}{v_0}}$. Fig. \ref{snapshots}(a)-(d) shows snapshots for $b^*=0.01$ and $c^*=0.1$. Other parameters are as in Fig. 2 in the main text, i.e. $\omega^*=3.0$ and $\alpha=4.0$).

\subsection{Robustness of network patterns against correlated noise acting on the center of mass coordinate}

Further we explore the influence of correlated random fields acting on the center of mass coordinate $\vec r^*_i$ of the particles. Therefore, we introduce three different Gaussian random fields $\phi_{\alpha,x}$, $\phi_{\alpha,y}$ and $\phi_{\alpha,\theta}$ with identical correlation exponent $\alpha$, leading to the following equations of motion:

\begin{align*}
 \dot{\vec r}_i^* &= \hat n_i(t^*) -\nabla_{r^*} U^*(r^*) + d^*\vec \phi_{\alpha}(\vec r^*,t^*) \\
 \dot{\theta}_i^* &= \omega^* + \kappa^*\sum_{j\in\partial_i}\sin(\theta_j^* - \theta_i^*) + a^*\phi_{\alpha,\theta}(\vec r_i^*,t^*).
\end{align*}

Here, $\vec \phi_{\alpha}= \left(\phi_{\alpha,x}, \phi_{\alpha,y} \right)$ and again, we neglect thermal fluctuations. Fig. \ref{snapshots}(e)-(h) shows snapshots for $d^*=0.2$, $\omega^*=3.0$ and $\alpha=4.0$. We clearly see that the network patterns which we discuss in the main text, persist in the presence of additional noise terms. When further increasing the strength of these additional noise terms, at some point, the network patterns cease to exist.

\begin{figure}
 \includegraphics[width=\textwidth]{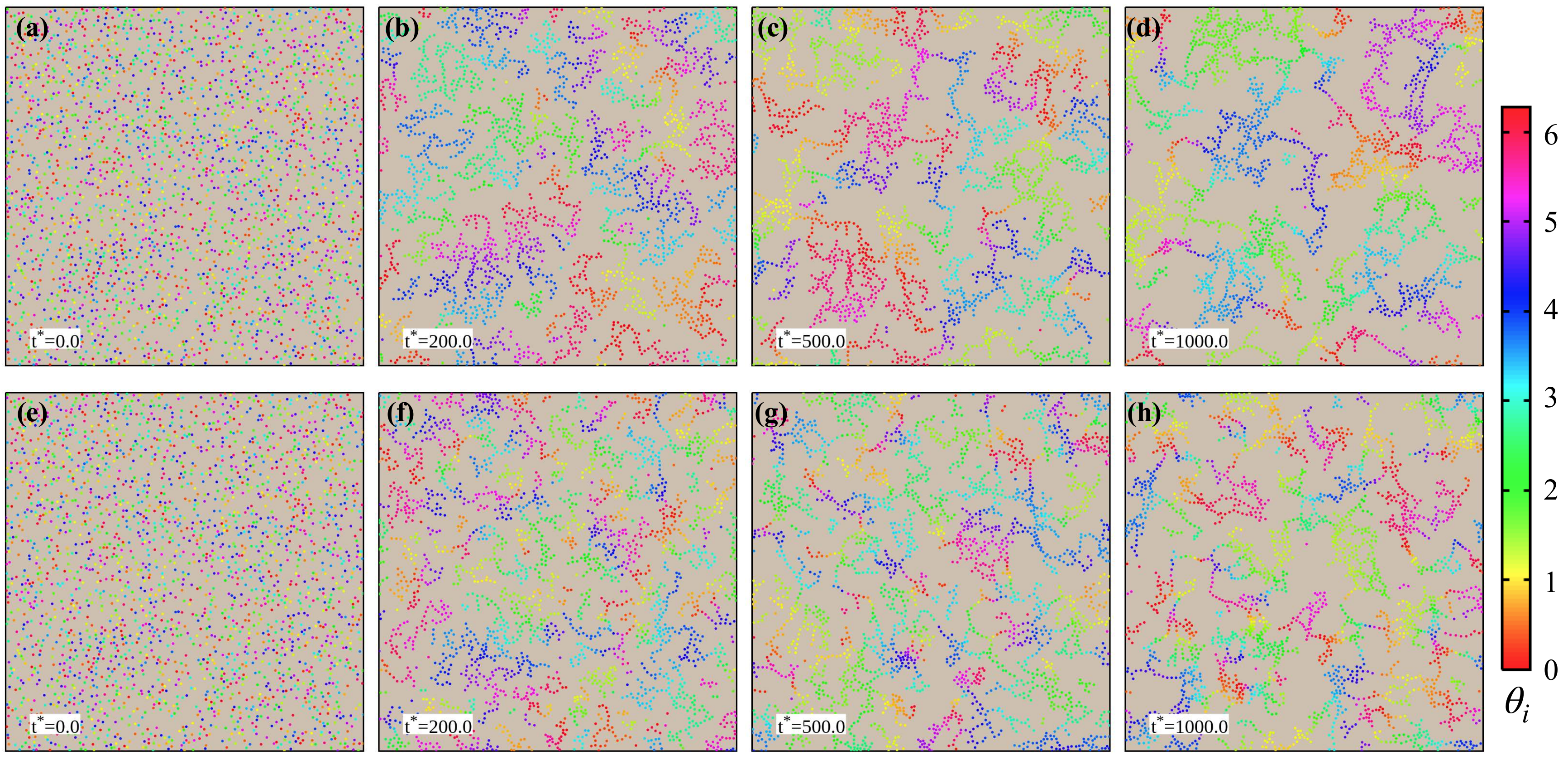}
 \caption{(a)-(d) Simulation snapshots with additional Gaussian white noise, where $b^*=0.01$ and $c^*=0.1$. (e)-(h) Snapshots of a simulation, where the spatially correlated but temporally uncorrelated random field also acts on the center of mass coordinates of the particles. Here, the amplitude of this additional field is $d^*=0.2$. As in the main text, the color coding indicates the particles orientation. Further parameters in both cases are: $\omega^*=3.0$, $\alpha=4.0$, $\kappa^*=1.0$, $\epsilon^*=1.0$, $a^*=1.0$ and $\Phi_0=0.1$.}\label{snapshots}
\end{figure}

\subsection{Role of short-range repulsions}

Finally, we explore the influence of volume exclusion interactions between the particles. Therefore, we perform simulations with $\epsilon^*=0$ and find, that also in that case, network patterns occur (see Fig. \ref{snapshots_vol_ex}). 

\begin{figure}
 \includegraphics[width=\textwidth]{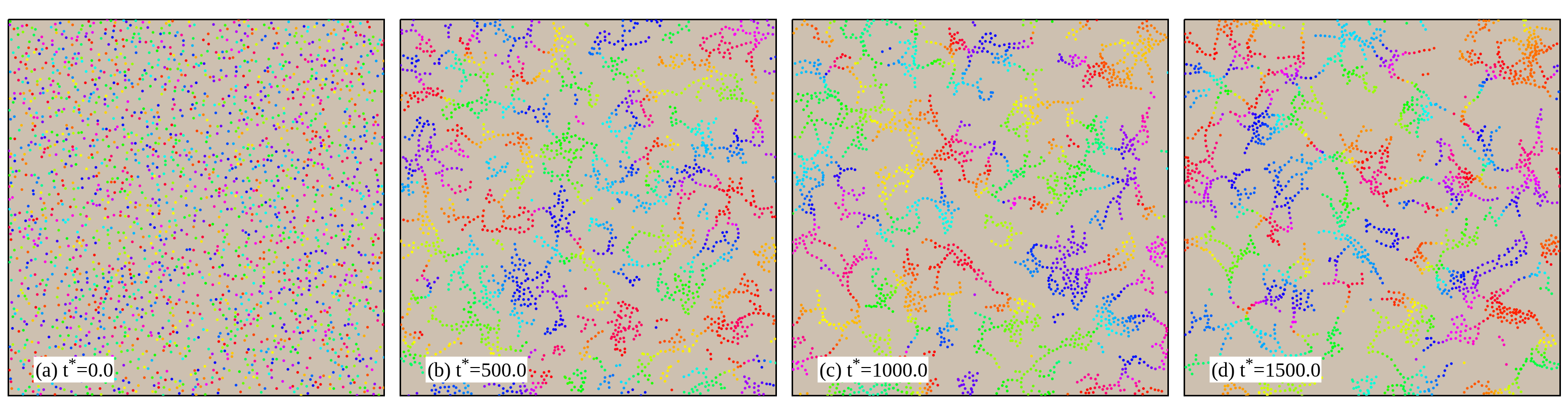}
 \caption{(a)-(d) Simulation snapshots without steric volume exclusion ($\epsilon^*=0$). Colors indicate particle orientations as in Fig. \ref{snapshots}. Further parameters are: $\omega^*=3.0$, $\alpha=4.0$, $\kappa^*=1.0$, $a^*=1.0$ and $\Phi_0=0.1$.}\label{snapshots_vol_ex}
\end{figure}

\section{Counting holes within a network pattern}

For calculating the number of holes in a network pattern (as shown in Fig. 3 in the main text), we divide the simulation box into bins with a width of two particle diameters and check for each bin, if it contains a particle or not. Then we use the bwlabel command of MATLAB (adopted to account for periodic boundary conditions) to find all neighboring bins, which are empty. Finally, we delete the largest hole which is essentially the environment of the pattern. Further we introduce a minimal hole size of two bins.

\section{Characteristic timescales}

The physical mechanism underlying the emergence of network patterns, as described in the main text, depends on the fact that the lifetime of dimers $\tau_\mathrm{c}$ is long compared to the average attachment time $\tau_{\mathrm{a}}$ for an additional particle to attach to a dimer. To estimate $\tau_{\mathrm{a}}$, in our simulations we first calculate the number of dimers during the simulation (see Fig. \ref{timescales}(a)) and then numerically calculate $\tau_{\mathrm{a}}$ via

\begin{equation*}
 \tau_{\mathrm{a}} = \frac{\int_0^{600} t^* N_{\mathrm{dim}}(t^*) \mathrm{d}t^*}{\int_0^{600} N_{\mathrm{dim}}(t^*) \mathrm{d}t^*} \approx 162.
\end{equation*}

Further, to determine the lifetime of dimers, we sequentially initialize $30$ dimers and calculate the distance $d_{\mathrm{dim}}$ of the two particles within the dimer. When the distance is larger than the range of the alignment interaction ($d_{\mathrm{dim}}>2$), the dimer counts as broken. We find that with uncorrelated noise, the typical lifetime of dimers is around $\tau_{\mathrm{uc}}\sim 38$, whereas for correlated noise, dimers are stable over the entire simulation (see Fig. \ref{timescales}(b)).
\begin{figure}[h!]
 \includegraphics[width=\textwidth]{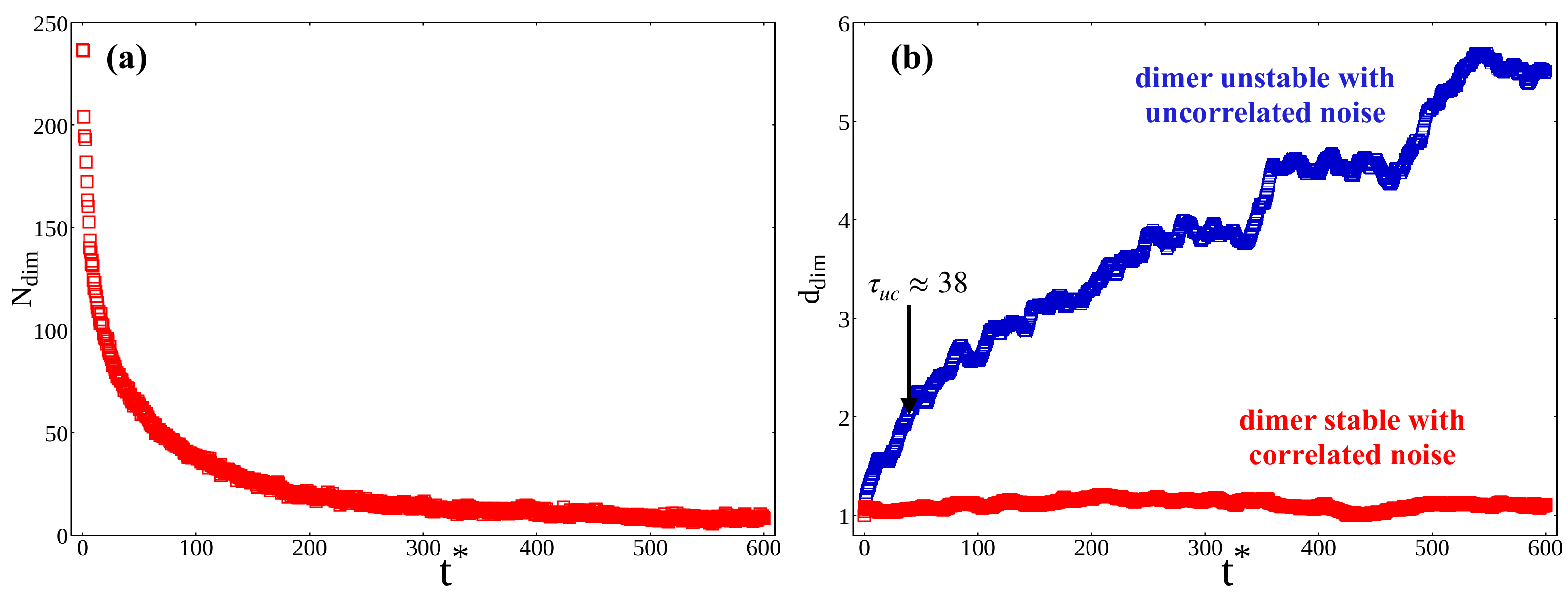}
 \caption{(a) Time evolution of the number of dimers $N_{\mathrm{dim}}$ in the system in the presence of correlated noise with $\alpha=4.0$, $\omega^*=3.0$, $\Phi_0=0.1$. (b) Averaged distance between two particles which initially build a dimer in the presence of uncorrelated (blue) and correlated (red) noise.}\label{timescales}
\end{figure}

\section{Movies}
In all movies the following parameters are fixed: $\Phi_0=0.1$, $\epsilon^*=1.0$ and $\kappa^*=1.0$.

\begin{itemize}
 \item Movie01-flocking: $\omega^*=0.0$ and $\alpha=4.0$, $a^*=1.0$;
 \item Movie02-clusters: $\omega^*=1.0$ and $\alpha=4.0$, $a^*=1.0$;
 \item Movie03-network-patterns: $\omega^*=3.0$ and $\alpha=4.0$, $a^*=1.0$;
 \item Movie04-without-noise: $\omega^*=3.0$ and $a^*=0.0$
\end{itemize}

\providecommand*{\mcitethebibliography}{\thebibliography}
\csname @ifundefined\endcsname{endmcitethebibliography}
{\let\endmcitethebibliography\endthebibliography}{}

\end{widetext}

\end{document}